\documentclass[submission,copyright,creativecommons]{eptcs}
\providecommand{\event}{Refinement 2018} % Name of the event you are submitting to
\usepackage{underscore}           % Only needed if you use pdflatex.

\usepackage{amssymb}
\usepackage{graphicx}
\usepackage{url}
\usepackage{booktabs}
\usepackage[space]{grffile}
\usepackage{color}
\definecolor{Light}{gray}{.95}
\usepackage[table]{colortbl}
\usepackage{amsmath}
\usepackage{yhmath}

\def\halfthinspace{\relax\ifmmode\mskip.5\thinmuskip\relax\else\kern.8888em\fi}

\newtheorem{definition}{Definition}
\newtheorem{proposition}{Proposition}
\newcommand {\probdist}{\theta}

\newcommand {\alphabet}{\alpha_a}
\newcommand {\Alphabet}{\alpha_A}
\newcommand {\numerics}{\nu}
\newcommand {\symbols}{\sigma}

\newcommand {\emptyrelation}{\phi}

\newcommand {\refinedby}{\sqsubseteq}

\newcommand {\tabeq}{\hspace*{0.2in}}
\newcommand {\tabrel}{\hspace*{0.60in}}

\newcommand {\nln}{@{}l@{}}
\newlength{\interligne}
\newcommand {\dpreuve}{\dimen123=\linewidth \dimen124=\linewidth
\advance\dimen123 by -20mm \advance\dimen124 by -5mm
\advance\dimen123 by -\mathindent \advance\dimen124 by -\mathindent
\setlength{\interligne}{\baselineskip}
\setlength{\baselineskip}{1.2\baselineskip}
    \begin{tabbing} 
    \hspace*{\mathindent}\= \hspace*{5mm}\= \kill 
    \+ \kill}
\newcommand {\fpreuve}{\end{tabbing}
    \setlength{\baselineskip}{\interligne}}

\newcommand {\dpreuveitem}{\begin{tabbing} 
    \hspace*{5mm}\= \hspace*{15mm}\= \kill}
\newcommand {\fpreuveitem}{\end{tabbing}}

\newcommand {\dspecitem}{\begin{tabbing} 
    \hspace*{5mm}\=\hspace*{5mm}\=\hspace*{5mm}\=
    \hspace*{5mm}\=\hspace*{5mm} \kill}
\newcommand {\fspecitem}{\end{tabbing}}

\def\[{\relax\ifmmode\@badmath\else\begin{trivlist}\item[]\leavevmode
  \hbox to\linewidth\bgroup$ \displaystyle
  \hskip\mathindent\bgroup\fi}
\def\]{\relax\ifmmode \egroup $\hfil \egroup \end{trivlist}\else \@badmath \fi}

\newcommand {\refines}{\sqsupseteq}

\newcommand {\dom}{\mbox{\it dom}}

\newcommand {\dtabin}{\begin{tabbing} 
    \hspace*{\mathindent}\= \kill \+ \kill}
\newcommand {\ftabin}{\end{tabbing}}
\newcommand {\dspec}{\begin{tabbing} 
    \hspace*{\mathindent}\= \hspace*{5mm}\=\hspace*{5mm}\=\hspace*{5mm}\=
                            \hspace*{5mm}\=\hspace*{5mm} \kill 
    \+ \kill}

\newcommand {\fspec}{\end{tabbing}}

\newlength {\longueurtop}
\newcommand {\initlongueurtop}{\setlength{\longueurtop}{\topsep}}
\newcommand {\topzero}{\setlength{\topsep}{0pt}}
\newcommand {\topdefaut}{\setlength{\topsep}{\longueurtop}}
\newcommand {\debuttab}{ \initlongueurtop \topzero \begin{tabbing} }
\newcommand {\fintab}{ \end{tabbing} \topdefaut }

\newcommand {\projection}{\Pi}

\title{Programming Without Refinement}
\author{Marwa Benabdelali \qquad\qquad Lamia Labed Jilani
\institute{ISG Management Institute\\ Bardo, Tunisia}
\email{\quad marwa.benabdelali@yahoo.com \quad\quad lamia.labed@isg.rnu.tn}
\and 
Wided Ghardallou
\institute{University of Kairouan\\ Kairouan, Tunisia}
\email{wided.ghardallou@gmail.com}
\and 
Ali Mili
\institute{New Jersey Institute of Technology\\ Newark, NJ, USA}
\email{mili@njit.edu}
}

\def\titlerunning{Programming Without Refinement}
\def\authorrunning{M. Benabdelali, L. Labed Jilani, W. Ghardallou \& A. Mili}

\begin{document}

%%\toctitle{Programming Without Refinement}
%%\tocauthor{Benabdelali, Jilani, Ghardallou, Mili}
\maketitle

\begin{abstract}
To derive a program for a given specification $R$ means to find an
artifact $P$ that satisfies two conditions:  $P$ is executable in some
programming language; and $P$ is correct with respect to $R$.
Refinement-based program derivation achieves this goal in a stepwise
manner by enhancing executability while preserving correctness until
we achieve complete executability.  In this paper, we argue that it is
possible to invert these properties, and to derive a program by enhancing
correctness while preserving executability (proceeding from one 
executable program to another) until we achieve absolute correctness.
Of course, this latter process is possible only if we know how
to enhance correctness.
\end{abstract}

%\subsection*{Keywords}
%
%Absolute correctness, relative correctness, program
%refinement,  correctness preservation,
%correctness enhancement, program derivation, program projection,
%deriving reliable programs.

\section{Introduction: Correctness Preservation vs Correctness Enhancement}

To derive a program from a specification $R$ means to find an
artifact $P$ that satisfies two conditions:  $P$ is executable in some
target programming language; and $P$ is correct with respect to $R$.
Refinement-based program derivation achieves this goal in a stepwise
manner by enhancing executability (substituting the
specification notation by programming notation) while maintaining
correctness until we achieve complete
executability.  In this paper we consider an orthogonal approach, where
these two properties are inverted:  We enhance correctness with respect
to $R$ while maintaining executability (all intermediate artifacts
are executable programs) until we achieve absolute correctness.
Figure \ref{twoprocessesfig} illustrates how these two iterative processes
differ.
\begin{figure}
\begin{center}
\begin{tabular}{|l|l|l|}
\hline\hline
% & & \\
Paradigm & Correctness Preservaation & Correctness Enhancement \\
% & & \\
 \hline\hline
\shortstack{Initial Condition} & $P=R$ & $P$ = {\tt abort} \\
\hline
Invariant Assertion & $P$ is correct &  $P$ is executable \\
\hline
Variant Function & Enhance Executability & Enhance Correctness \\
\hline
Exit Condition & $P$ is Executable & $P$ is Correct \\
\hline\hline
\end{tabular}
\end{center}
\caption{\label{twoprocessesfig}Orthogonal Derivation Processes}
\end{figure}

Program derivation by correctness enhancement was introduced in
\cite{oslo}.  In this paper we build on the discussions of
\cite{oslo} by:  
Considering more sample examples of program
derivation by correctness enhancement (this is the subject of section
\ref{samplesection}); 
in light of our experience with these sample
examples, sketching the first outlines of a methodology
of correctness enhancement;
considering the concept of {\em projection}, and its impact
on the discipline of program derivation by correctness enhancement;
using empirical evidence to analyze the evolution of program
reliability through the correctness enhancement process.

In section \ref{backgroundsection} we introduce some elements of
relational mathematics, then we discuss the basic concepts that we
need in this paper.
In section
\ref{samplesection} we present a number of sample program derivations
by correctness enhancement, and in section \ref{reliabilitysection} we
analyze the reliability growth of the programs generated in each example, using
a simple experimental set-up.  
In section \ref{takingstocksection}
we take stock of the experience gained through the
examples of section \ref{samplesection} to sketch the outlines
of a programming methodology that is adapted to correctness
enhancement, and discuss the contrast between correctness
enhancement and related properties.  Finally, in section
\ref{conclusionsection} we summarize and assess our findings,
and sketch directions of future
research.

\section{Mathematics for Correctness}
\label{abscorsect}
\label{backgroundsection}

\subsection{Relational Mathematics}
\label{mathbackgroundsect}

We assume the reader familiar with elementary relational mathematics
\cite{brink}, and will merely present some
definitions and notations.
We represent sets in a program-like notation by writing variable names
and associated data types; if we write $S$ as:
{\tt \{x: X; y: Y;\}},
then we mean to let $S$ be the cartesian product $S=X\times Y$; elements of $S$
are denoted by $s$ and the $X$- (resp. $Y$-) component of $s$
is denoted by $x(s)$ (resp. $y(s)$).  When no ambiguity arises, we may
write $x$ for $x(s)$, and $x'$ for $x(s')$, etc.  A {\em relation} $R$ on
set $S$ is a subset of $S\times S$.  Special relations on $S$ include
the {\em universal relation} $L=S\times S$, the identity relation $I=\{(s,s)|
s\in S\}$ and the empty relation $\emptyrelation=\{\}$.  Operations on relations
include the set theoretic operations of union, intersection, difference
and complement; they also include the {\em converse} of a relation $R$
defined by $\widehat{R}=\{(s,s')| (s',s)\in R\}$, the {\em domain} of a 
relation defined by $\dom(R)=\{s| \exists s': (s,s')\in R\}$, and the product
of two relations $R$ and $R'$ defined by: $R\circ R'=\{(s,s')|
\exists s'': (s,s'')\in R\wedge (s'',s')\in R'\}$; when no ambiguity
arises, we may write $RR'$ for $R\circ R'$.

A relation $R$ is said to be {\em reflexive} if and only if $I\subseteq R$,
{\em symmetric} if and only if $R=\widehat{R}$, {\em antisymmetric} 
if and only if $R\cap\widehat{R}
\subseteq I$, {\em asymmetric} if and only if $R\cap\widehat{R}=\emptyrelation$ and
{\em transitive} if and only if $RR\subseteq R$.  A relation $R$ is said to be
{\em total} if and only if $I\subseteq R\widehat{R}$ and {\em deterministic}
if and only if $\widehat{R}R\subseteq I$ (we then say that $R$ is a 
{\em function}).  A relation $R$ is said to be a 
{\em vector} if and only if $RL=R$; vectors have the form $R=A\times S$
for some subset $A$ of $S$; we use them as relational representations of
sets.  In particular, note that $RL$ can be written as $\dom(R)\times S$;
we use it as a representation of the domain of $R$.

\subsection{Program Semantics}

Given a program {\tt p} on space $S$,
we define the {\em function} of {\tt p} (denoted by $P$)
as the set of pairs
$(s,s')$ such that if program {\tt p} starts execution in state $s$
it terminates in state $s'$; when no ambiguity arises, we may refer
to a program and its function by the same name, $P$.
\begin{definition}
\label{refinedef}
Given two relations $R$ and $R'$, we say that $R'$ {\em refines} $R$
(abbrev:  $R'\refines R$ or $R\refinedby R'$)
if and only if %$RL\cap R'L\cap (R\cup R')=R$.
$RL\subseteq R'L\wedge RL\cap R'\subseteq R$.
\end{definition}
This is the relational form of the usual interpretation of
refinement as having a weaker precondition and a stronger postcondition.

\begin{definition}
\label{abscorrectdef}
A program {\tt p} on space $S$ is said to be {\em correct} with respect to
specification $R$ on $S$ if and only if its function $P$ refines $R$.
\end{definition}
This definition is identical (modulo differences of notation) to
traditional definitions of total correctness
\cite{back,gries,hehnerec,manna}.

\subsection{Relative Correctness}
\label{relcorsect}

\begin{definition}
\label{relcordetdef}
Due to \cite{ramics}.  
Given a specification $R$ and two deterministic programs $P$ and $P'$, we say that
$P'$ is {\em more-correct}
(resp. {\em strictly more-correct}) than $P$ with respect to $R$,
denoted as $P'\refines_R P$ (resp. $P'\sqsupset_R P$) if and only if
$(R\cap P')L\supseteq (R\cap P)L$ (resp. $(R\cap P')L\supset (R\cap P)L$).
\end{definition}
To contrast relative correctness with correctness,
we may refer to the latter as
{\em absolute correctness}.  We refer to $(R\cap P)L$ (or $\dom(R
\cap P)$) as the
{\em competence domain} of $P$ with respect to $R$.
See Figure \ref{relcorfig} for an illustration of relative correctness.
We have: $(R\cap P)=\{(1,2),(2,3)\}$, hence $(R\cap P)L=\{1,2\}\times S$.  On the
other hand, $(R\cap P')=\{(1,0),(2,1),(3,2)\}$, hence $(R\cap P')L=\{1,2,3\}\times S$.

\begin{figure*}
\thicklines
\setlength{\unitlength}{0.022in}
\begin{center}
\begin{picture}(170,35)

\put(0,0) {\makebox(0,0){3}}
\put(0,10){\makebox(0,0){2}}
\put(0,20){\makebox(0,0){1}}
\put(0,30){\makebox(0,0){0}}

\put(50,0) {\makebox(0,0){3}}
\put(50,10){\makebox(0,0){2}}
\put(50,20){\makebox(0,0){1}}
\put(50,30){\makebox(0,0){0}}

\put(60,0) {\makebox(0,0){3}}
\put(60,10){\makebox(0,0){2}}
\put(60,20){\makebox(0,0){1}}
\put(60,30){\makebox(0,0){0}}

\put(110,0) {\makebox(0,0){3}}
\put(110,10){\makebox(0,0){2}}
\put(110,20){\makebox(0,0){1}}
\put(110,30){\makebox(0,0){0}}

\put(120,0) {\makebox(0,0){3}}
\put(120,10){\makebox(0,0){2}}
\put(120,20){\makebox(0,0){1}}
\put(120,30){\makebox(0,0){0}}

\put(170,0) {\makebox(0,0){3}}
\put(170,10){\makebox(0,0){2}}
\put(170,20){\makebox(0,0){1}}
\put(170,30){\makebox(0,0){0}}

%R
\put(5,30){\vector(1,0){40}}
\put(5,0){\vector(1,0){40}}
\put(5,0){\vector(4,1){40}}
\put(5,20){\vector(4,1){40}}
\put(5,20){\vector(4,-1){40}}
\put(5,10){\vector(4,1){40}}
\put(5,10){\vector(4,-1){40}}

%P
\put(65,30){\vector(4,-1){40}}
\put(65,20){\vector(4,-1){40}}
\put(65,10){\vector(4,-1){40}}

%P'
\put(125,20){\vector(4,1){40}}
\put(125,10){\vector(4,1){40}}
\put(125,0){\vector(4,1){40}}

{\scriptsize
\put(25,35){\makebox(0,0){$R$}}
\put(85,35){\makebox(0,0){$P$}}
\put(145,35){\makebox(0,0){$P'$}}}

\put(60,15){\oval(4,20)}
\put(120,10){\oval(4,30)}

\end{picture}
\end{center}
\caption{\label{relcorfig}
$P'\refines_R P$, Deterministic Programs}
\end{figure*}

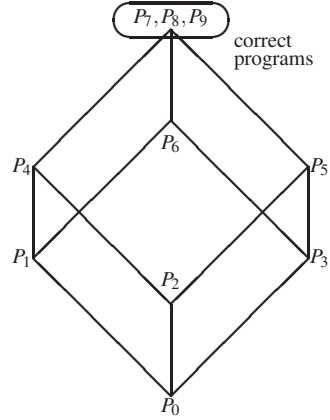
\begin{figure}
\thicklines
\setlength{\unitlength}{0.012in}
\begin{center}
\begin{picture}(130,170)

\put(65,5){\line(1,1){60}}
\put(65,5){\line(-1,1){60}}
\put(65,5){\line(0,1){40}}

\put(65,125){\line(1,-1){60}}
\put(65,125){\line(-1,-1){60}}
\put(65,125){\line(0,1){40}}

\put(5,105){\line(0,-1){40}}
\put(5,105){\line(1,-1){60}}
\put(5,105){\line(1,1){60}}

\put(125,105){\line(0,-1){40}}
\put(125,105){\line(-1,-1){60}}
\put(125,105){\line(-1,1){60}}

{\scriptsize
\put(65,0){\makebox(0,0){$P_0$}}
\put(130,65){\makebox(0,0){$P_3$}}
\put(130,105){\makebox(0,0){$P_5$}}
\put(65,55){\makebox(0,0){$P_2$}}
\put(65,115){\makebox(0,0){$P_6$}}
\put(65,170){\makebox(0,0){$P_7, P_8, P_9$}}
\put(0,65){\makebox(0,0){$P_1$}}
\put(0,105){\makebox(0,0){$P_4$}}
\put(110,155){\makebox(0,0){\shortstack[l]{correct\\ programs}}}}

\put(65,165){\makebox(0,0){\oval(50,15)}}

\end{picture}
\end{center}
\caption{\label{candidatefig}
Ordering Candidate Programs by Relative Correctness}
\end{figure}

How do we know that our definition is any good?
In \cite{ramics}, we find that relative correctness satisfies the
following properties:
\begin{itemize}
\item {\em Ordering Properties}.
Relative correctness is reflexive and transitive, but not antisymmetric.
Two programs $P$ and $P'$ may be equally correct yet 
distinct.
\item {\em Relative Correctness and Absolute Correctness}.
A (absolutely) correct program is more-correct than (or as correct as)
any candidate program. % We write this as:
%$$\boxed{P'\refines R\Leftrightarrow (\forall P, P'\refines_R P).}$$
A deterministic program $P$ is (absolutely) correct 
with respect to $R$ if and only if its
competence domain is $\dom(R)$.
\item {\em Relative Correctness and Reliability}.
The reliability of a program $P$ on space $S$ 
is defined with respect to a specification $R$ on $S$
and a discrete probability distribution $\probdist()$ on $\dom(R)$.  We
measure it by the probability that the execution of $P$ on a random
state $s$ of $\dom(R)$ selected according to $\probdist()$ terminates
successfully in a state $s'$ such that $(s,s')\in R$; in other words,
%the reliability of a program $P$ with respect to a specification $R$
%and a probability distribution $\probdist()$ 
it is the probability
that a randomly selected element of $\dom(R)$ following the probability
distribution $\probdist()$ falls within the competence domain of $P$
with respect to $R$. % Accordingly,
%the reliability of program $P$ can be written as:
%$$\reliability_R^{\probdist}(P)=\sum_{s\in\dom(R\cap P)} \probdist(s).$$
From this definition, and from definition \ref{relcordetdef}, we
infer that if $P'$ is more-correct than $P$, then $P'$ is more
reliable than $P$ for any probability distribution $\probdist()$.
%Sufficiency can be proven easily:  if the competence domain of $P'$
%is not a superset of that of $P$, then there exists an element of the
%competence domain of $P$ that is outside the competence domain of $P'$;
%we can easily define a probability distribution $\probdist$ for
%which $P$ has higher reliability than $P'$.
%Hence we write:
%$$\boxed{P'\refines_R P\Leftrightarrow (\forall\probdist:
%\reliability_R^{\probdist}(P')\geq\reliability_R^{\probdist}(P)).}$$
\item {\em Relative Correctness and Refinement}.  Program $P'$ refines
program $P$ if and only if $P'$ is more-correct than $P$ with respect to
{\em any} specification $R$.%  We write this property as:
%$$\boxed{P'\refines P\Leftrightarrow (\forall R:  P'\refines_R P).}$$
%Necessity stems from the fact that for deterministic programs $P$ and $P'$,
%$P'\refines P$ is equivalent to $P'\supseteq P$.  Sufficiency can be
%inferred by taking $R=P$.
\end{itemize}
For illustration, we present below a specification and ten
programs, ranked by relative correctness, as shown
in Figure \ref{candidatefig};
correct programs are shown at the top of the graph.
We let $R$ be the specification defined on space
$S=nat$ by:  
$R=\{(s,s')| s^2\leq s'\leq s^3\},$
and we consider the following programs, where with each program
we indicate its function, and its competence domain:
\begin{itemize}
\item [{\tt p0}:] {\tt \{abort\}}. $P_0=\emptyrelation$. $\textit{CD}_0=\emptyset$.
\item [{\tt p1}:] {\tt \{s=0;\}}.  $P_1=\{(s,s')| s'=0\}$.  $\textit{CD}_1=\{0\}$.
\item [{\tt p2}:] {\tt \{s=1;\}}.  $P_2=\{(s,s')| s'=1\}$.  $\textit{CD}_2=\{1\}$.
\item [{\tt p3}:] {\tt \{s=2*s**3-8;\}}.  $P_3=\{(s,s')| s'=2s^3-8\}$.  $\textit{CD}_3=\{2\}$.
\item [{\tt p4}:] {\tt \{skip;\}}.  $P_4=I$.  $\textit{CD}_4=\{0,1\}$.
\item [{\tt p5}:] {\tt \{s=2*s**3-3*s**2+2;\}}.  $P_5=
\{(s,s')| s'=2s^3-3s^2+2\}$.  $\textit{CD}_5=\{1,2\}$.
\item [{\tt p6}:] {\tt \{s=s**4-5*s;\}}.  $P_6=
\{(s,s')| s'=s^4-5s\}$.  $\textit{CD}_6=\{0,2\}$.
\item [{\tt p7}:] {\tt \{s=s**2;\}}.  $P_7=\{(s,s')| s'=s^2\}$.  $\textit{CD}_7=S$.
\item [{\tt p8}:] {\tt \{s=s**3;\}}.  $P_8=\{(s,s')| s'=s^3\}$.  $\textit{CD}_8=S$.
\item [{\tt p9}:] {\tt \{s=(s**2+s**3)/2;\}}.  $P_9=\{(s,s')| 
s'=\frac{s^2+s^3}{2}\}$.  $\textit{CD}_9=S$.
\end{itemize}

The following definition
applies to non-deterministic programs.
\begin{definition}
\label{relcornddef}
Due to \cite{ramicsdesh}.
Given a specification $R$ and two programs $P$ and $P'$.  We say that $P'$ is
{\em more-correct} than $P$ with respect to $R$ if and only if:
$$(R\cap P')L\supseteq (R\cap P)L\wedge(R\cap P)L\cap\overline{R}\cap P'\subseteq
P.$$
%Also, we say that $P'$ is {\em strictly more-correct} than $P$ with respect 
%to $R$ if and only if 
%at least one of the inequalities in this definition is strict.
\end{definition}

\subsection{Program Projection}
\label{projectionsection}

We consider a space $S$ defined by two integer variables $x$ and $y$, and
we let $R$ be the following specification:  $R=\{(s,s')| x'=x+y\}$.  We let
{\tt p} be the following candidate program:\\ \tabeq
{\tt \{while (y!=0) \{x=x+1; y=y-1;\}\}}.\\
The function of {\tt p} is:
$P=\{(s,s')| y\geq 0\wedge x'=x+y\wedge y'=0\}$.\\
When we consider this function, we find that it has clauses (e.g. $x'=x+y$)
that are mandated
by the specification $R$ and clauses (e.g. $y'=0$) that are not mandated by $R$
but stem instead from the design of $P$.
In \cite{desharnais}, we introduce an operator $\projection_R(P)$, called the
{\em projection of $P$ over $R$}, which represents the functionality of $P$
that is mandated by $R$.  In the example above, we want 
the projection of $P$ over $R$ to be:
$$\projection_R(P)=\{(s,s')| y\geq 0\wedge x'=x+y\}.$$
Indeed, $P$ delivers
$(y'=0)$ but $R$ does not require it; and $R$ mandates $(x'=x+y)$ for
negative $y$ but $P$ does not deliver it. 
\begin{definition}  Due to \cite{desharnais}.
Given a specification $R$ on space $S$ and a program $P$ on $S$, the
{\em projection of $P$ over $R$} is the relation denoted by $\projection_R(P)$
and defined as $(R\cap P)L\cap(R\cup P)$.
\end{definition}
The importance of projections is reflected in the following
proposition (\cite{desharnais}).
\begin{proposition}
\label{relcorrefineprop}
Given a specification $R$ and two programs $P$ and $P'$, $P'$ is more-correct
than $P$ with respect to $R$ if and only if $\projection_R(P')$ refines
$\projection_R(P)$.
\end{proposition}
Since the projection
of a program on a specification reflects the functionality of the program
that is relevant to the specification, it is only normal that it be
the only part of $P$ that
determines relative correctness with respect to the specification.

\section{Sample Examples}
\label{samplesection}

\subsection{Fermat Decomposition}

This example, due to \cite{oslo}, uses a specification due to 
\cite{dromey}. 
We let space $S$ be defined by natural variables $n$, $x$ and $y$, and we
let specification $R$ be defined as:\\
\tabeq 
$R = \{(s,s')| n=x'^2-y'^2\wedge 0\leq y'\leq x'\}.$\\
The domain of $R$ is the set of states $s$ such that $n(s)$ 
is either odd or a multiple 
of 4.
Hence we write:
$RL=\{(s,s')| n \mod 2=1\lor n\mod 4=0\}.$
Whereas Dromey \cite{dromey}
presents a sequence of designs that are increasingly more concrete,
we present a
sequence of programs that are increasingly more-correct.
Starting from the initial program $P_0=${\tt abort}, we resolve to let
the next program $P_1$ find the
required factorization for
$y'=0$:
\begin{verbatim}
void p1()
   {nat n, x, y; x=0; y=0;    
    {nat r; r=0; while (r<n) {r=r+2*x+1; x=x+1;}}}
\end{verbatim}
The function of this program is:\\
\tabeq $P_1=\{(s,s')| n'=n\wedge y'=0\wedge x'=\lceil\sqrt{n}~\rceil\}.$\\
Whence we compute the competence domain of $P_1$ with respect to $R$:\\
\tabeq $(R\cap P_1)L
=\{(s,s')| \exists x'':  n=x''^2\}$.\\
In other words, $P_1$ satisfies specification $R$ whenever $n$ is a perfect square.
We now consider the case where $r$ exceeds $n$ by a perfect square, making it
possible to fill the difference with $y^2$:
\begin{verbatim}
void p2() {nat n, x, y;   //  input/output variables
    x=0; y=0; {nat r; r=0; while (r<n) {r=r+2*x+1; x=x+1;}
    if (r>n) {while (r>n) {r=r-2*y-1; y=y+1;}}}}
\end{verbatim}
The function of this program is:\\
\tabeq
$P_2=\{(s,s')| n'=n\wedge x'=\lceil\sqrt{n}~\rceil\wedge y'^2=x'^2-n\wedge y'\geq 0\}.$\\
The competence domain of $P_2$ with respect to $R$ is:\\
\tabeq $(R\cap P_2)\circ L
=\{(s,s')| \exists y'':
y''^2=\lceil\sqrt{n}~\rceil^2-n\}$.\\
The competence domain of $P_2$ is the set of states $s$ such that
the difference between $n(s)$ and the
square of the ceiling of the square root of $n(s)$ is a perfect square.
This is a superset of the competence domain of 
$P_1$, hence $P_2$ is more-correct than $P_1$.
The next program is derived from $P_2$ by resolving that if the ceiling of
the integer square root of $n$ does not exceed $n$ by a perfect square, then
we try the next perfect square, etc. 
We know that this process converges for any state $s$ for which
$n(s)$ is odd or a multiple of 4.  Hence,
\begin{verbatim}
void p3() {nat n, x, y;   //  input/output variables
    {nat r; x=0; r=0; while (r<n) {r=r+2*x+1; x=x+1;}
    while (r>n)  {int rsave; y=0; rsave=r;
        while (r>n) {r=r-2*y-1; y=y+1;}
        if (r<n) {r=rsave+2*x+1; x=x+1;}}}}
\end{verbatim}
If we let
$\mu(n)$ be the smallest number whose 
square exceeds $n$ by a perfect square, we write the function
of $P_3$ as follows:\\
\tabeq 
$P_3=\{(s,s')| n'=n\wedge x'=\mu(n)\wedge y'=\sqrt{\mu(n)^2-n}\}.$\\
We compute the competence domain of $P_3$ with respect to $R$ and we find:
$(R\cap P_3)\circ L=RL$.
Hence $P_3$ is correct with respect to $R$
hence it is more-correct than $P_2$ with respect to $R$.  Hence we do have:\\
\tabeq
$P_0\refinedby_R P_1\refinedby_R P_2\refinedby_R P_3.$\\
Furthermore, we find that $P_3$ is correct with respect to $R$.

\subsection{The Ceiling of the Square Root}

This example, due to Reinfelds \cite{reinfelds}, consists in computing
the non-negative integer square root of a non-negative integer $n$; its space
is defined by variables $n$ and $x$ of type integer, and its specification is
written as:
$$R=\{(s,s')| x'^2\leq n<(x'+1)^2 \wedge x'\geq 0\}.$$
In \cite{reinfelds} Reinfelds offers two solutions to this problem;
both solutions focus on the derivation of a while loop, and both start by
deriving a loop invariant from the post-condition.  
We let the first program be
{\tt p0:  abort},
whose competence domain ($CD_0$)
is the empty set, and for
the next program, we choose:\\
\tabeq {\tt p1:  \{int x, n; x=0;\}}.\\
The function of this program and its competence domain are given as: \\ 
\tabeq $P_1=\{(s,s')| x'=0\wedge n'=n\}$\\
\tabeq $CD_1=(R\cap P_1)L=\{(s,s')|0\leq n<1\wedge x'=0\wedge n'=n\}L=\{(s,s')|n=0\}$.\\
This program satisfies the specification only for $n=0$.
For the next program, we want to satisfy $R$ whenever $n$ is a perfect
square.\\
\tabeq {\tt p2: \{int x,n;x=0; \{int x2=0; while(x2<n)\{x2=x2+2*x+1;x=x+1;\}\}\}}.\\
We compute the function of the while loop using invariant relations, as we discuss in
\cite{miliscp},
and we find the following function and competence domain for $P_2$:\\
\tabeq $P_2=\{(s,s')| (x'-1)^2<n\leq x'^2\wedge x'\geq 0\wedge n'=n\}$.\\
\tabeq $CD_2=%(R\cap P_2)L$\\
%\tabeq $=\{(s,s')| x'^2\leq n<(x'+1)^2\wedge x'\geq 0\wedge
%(x'-1)^2<n\leq x'^2\wedge n'=n\}L$\\
%\tabeq $=\{(s,s')| n=x'^2\wedge x'\geq 0\wedge n'=n\}L
\{(s,s')|\exists x': n=x'^2\}$,\\
which means that $n$ is a perfect square.  
For the fourth program, we want to satisfy specification $R$ even when
$n$ is not a perfect square.  We consider the following program:\\
\tabeq {\tt p3: \{int x, n; x=0;}\\
\tabeq \tabeq \tabeq 
{\tt \{int x2=0; while(x2<n)\{x2=x2+2*x+1;x=x+1;\} if (x2>n) \{x=x-1;\}\}\}}\\
The function of program $P_3$ can be obtained from that of $P_2$ by multiplying
it on the right by the function of the {\tt if-then} statement, which is:\\
\tabeq $F=\{(s,s')| x^2>n\wedge x'=x-1\wedge n'=n\}\cup\{(s,s')| x^2=n\wedge s'=s\}$.\\
By computing the product, then simplifying the terms, we find:\\
%\tabeq $P_3=\{(s,s')| n=x'^2\wedge x'\geq 0\wedge n'=n\}$\\
%\tabrel $\cup\{(s,s')| x'^2<n<(x'+1)^2\wedge x'\geq 0\wedge n''=n\}.$\\
%Merging the terms of the union, we find:\\
\tabeq $P_3=\{(s,s')| x'^2\leq n<(x'+1)^2\wedge x'\geq 0\wedge n''=n\}.$\\
The competence domain of $P_3$ is:
%\tabeq $CD_3=(R\cap P_3)L=\{(s,s')|x'^2\leq n<(x'+1)^2\wedge x'\geq 0\wedge n''=n\}L$\\ 
$CD_3=\{(s,s')| n\geq 0\}$.
This is equal to the domain of $R$, hence $P_3$ is correct with respect to $R$.
%Because this competence domain is a superset of $CD_2$, we infer that $P_3$
%is more-correct than $P_2$ with respect to $R$.  In addition, because the
%competence domain of $P_3$ is actually equal to $RL$, we conclude that $P_3$
%is absolutely correct with respect to $R$.

\subsection{Analyzing a String}

This example is due to \cite{gonzalezsanchez}, and aims to scan
a sequence $q$ and count the number of letters, digits and other
symbols.
We let $S$ be the space defined by a variable $q$ of type {\em string}
and integer variables $let$, $dig$, and $other$;  and we let $R$ be defined as:\\
\tabeq
$R=\{(s,s')| q\in list
\langle\alpha_A \cup \alpha_a \cup \vartheta \cup \sigma\rangle \wedge $\\
\tabrel $
let'=\#_a(q)+\#_A(q) \wedge dig'=\#_\vartheta(q) \wedge other'=\#_\sigma(q)\}$\\
where $list\langle T\rangle$ denotes the set of lists of elements of type $T$, 
and 
$\#_A$, $\#_a$, $\#_\vartheta$ and $\#_\sigma$ denote the functions that to each 
list $l$ assign (respectively) the number of upper case alphabetic 
characters, lower case alphabetic characters, numeric digits and symbols.
We generate the following programs:
\begin{itemize}
\item[{\tt p0:}] {\tt\{abort\}}.  $CD_0=\emptyrelation$.
\item[{\tt p1:}] 
\begin{verbatim}
{i=0; let=0; dig=0; other=0; l=strlen(q);
while (i<l) {c=q[i];i++; if ('A'<=c && 'Z'>=c) let+=1;}}
\end{verbatim}
$CD_1 = \{(s,s')| q\in list\langle\Alphabet\rangle\}$. 
\item[{\tt p2:}] 
\begin{verbatim}
{i=0; let=0; dig=0; other=0; l=strlen(q);
while (i<l) {c = q[i]; i++; 
   if ('A'<=c && 'Z'>=c) let+=1;     
   else if ('a'<=c && 'z'>=c) let+=1;}}
\end{verbatim}
$CD_2 = \{(s,s')| q\in list\langle\Alphabet\cup\alphabet\rangle\}$. 
\item[{\tt p3:}] 
\begin{verbatim}
{i=0; let=0; dig=0; other=0; l=strlen(q);
while (i<l) {c = q[i]; i++;                                                   
   if ('A'<=c && 'Z'>=c) let+=1;     
   else  if ('a'<=c && 'z'>=c) let+=1; 
   else  if ('0'<=c && '9'>=c) dig+=1;}}
\end{verbatim}
$CD_3 = \{(s,s')| q\in list\langle\Alphabet\cup\alphabet\cup
\numerics\rangle\}$. 
\item[{\tt p4:}] 
\begin{verbatim}
{i=0; let=0; dig=0; other=0; l=strlen(q);
while (i<l) {c = q[i]; i++; 
   if ('A'<=c && 'Z'>=c) let+=1;     
   else if ('a'<=c && 'z'>=c) let+=1;
   else if ('0'<=c && '9'>=c) dig+=1;
   else other+=1;}}
\end{verbatim}
$CD_4 = \{(s,s')| q\in list\langle\Alphabet\cup\alphabet\cup
\numerics\cup\symbols\rangle\}$. 
\end{itemize}
Since $CD_0\subseteq CD_1\subseteq CD_2\subseteq CD_3\subseteq CD_4$, 
we do have $P_0\refinedby_R P_1\refinedby_R P_2\refinedby_R P_3
\refinedby_R P_4$; also, we find $CD_4=RL$,
hence $P_4$ is absolutely correct
with respect to $R$.

\subsection{Word Wrap}

The specification of this problem is borrowed from \cite{meyer,perelman};
for the sake of readability and brevity,
we present the English text of
the specification (due to \cite{mili}), but not the relational representation.

"The program accepts as input a finite sequence of characters and produces as
output a sequence of characters satisfying the following conditions:
\begin{itemize}
\item If the input sequence contains MaxPos+1 consecutive non-break characters
then a boolean flag (longWord) is set to true.
\item Else,
\begin{itemize}
\item LongWord is set of false.
\item All the words of the input appear in the output, 
in the same order, and all the words of 
the output appear in the input.
\item Furthermore, the output must must satisfy the following conditions:
\begin{itemize}
\item It contains no leading or trailing breaks, nor consecutive breaks, where
a break is a blank, or a newline or the end-of-file.
\item Any sequence of MaxPos+1 consecutive characters includes a newline.
\item Any subsequence made up of no more than MaxPos characters and embedded
between the head of the sequence or a new line on the left, and the tail of
the sequence or a break on the right contains no newline."
\end{itemize}
\end{itemize}
\end{itemize}

Due to space limitations, we do not compute the function of each program
in our sequence of solutions, but content ourselves with presenting their
competence domains.
We generate the following sequence of programs for this specification, starting
from {\tt p0:  abort}, whose competence domain is empty.
The first program merely echos the input to the output; we generate it to pin down
the mechanics of file transfer in C++.
\begin{verbatim}
p1:  #include <fstream> using namespace std;
const int MaxPos = 80; const char blank = ' ';
ifstream inpstr;  ofstream outpstr;
int main ()
   {inpstr.open("inp1.dat"); outpstr.open("outp1.dat");
    char c;  c=inpstr.get();
    while (!inpstr.eof()) {outpstr << c; c=inpstr.get();}
    inpstr.close(); outpstr.close();}
\end{verbatim}
The competence domain ($CD_1$) of this program is the set of input sequences that are
not longer than MaxPos, have no newlines, no leading or trailing blanks, and
single blanks between words.
The second program assumes that the input contains no newlines, and
merely removes leading and trailing blanks, as well as extra blanks
between words.
\begin{verbatim}
p2:  #include <fstream>
#include <string> using namespace std;
const int MaxPos = 80; const char blank = ' ';
const string emptyword="";
ifstream inpstr; ofstream outpstr; char c;  string word;
void skipblanks();
void echoword();
int main ()
   {inpstr.open("inp2.dat"); outpstr.open("outp2.dat");
    c=inpstr.get(); skipblanks();
    while (!inpstr.eof()) {echoword();}
    inpstr.close(); outpstr.close();}
void skipblanks()
   {while ((!inpstr.eof()) && (c==blank)) {c=inpstr.get();};}
void echoword ()
   {bool leadingblanks; leadingblanks=(c==blank); skipblanks();
    string word; word=emptyword; 
    while ((!inpstr.eof())&&(c!=blank)) {word+=c;c=inpstr.get();}
    if (word.length()>0) {if (leadingblanks) 
    {outpstr << blank << word;} else {outpstr << word;}}}
\end{verbatim}
The competence domain ($CD_2$) of this program is the set of sequences whose
compacted version (when extra blanks are removed) is not longer than MaxPos,
and have no newlines.
The third program removes newlines in addition to extra spaces throughout
the data stream.
\begin{verbatim}
p3:  //  same as p2, except:
const char lf = '\n'; const char cr = '\r'; 
void skipblanks()
   {while ((!inpstr.eof())&&((c==blank)||(c==lf)||(c==cr))) 
      {c=inpstr.get();}}
void echoword ()
   {bool leadingblanks; 
    leadingblanks=((c==blank)||(c==lf)||(c==cr)); skipblanks();
    string word; word=emptyword; 
    while ((!inpstr.eof()) && (c!=blank) && (c!=lf) && (c!=cr)) 
       {word +=c; c=inpstr.get();}
    if (word.length()>0) {if (leadingblanks) 
       {outpstr << blank << word;} else {outpstr << word;}}}
\end{verbatim}
The competence domain ($CD_3$) of this program is the set of sequences whose
compacted version (when blanks and newlines are removed) is not longer than
MaxPos.
The fourth program places newlines at the appropriate places in the output
stream.
\begin{verbatim}
p4:
//  same as p3 except:
int linelen;
void echoword ()
   {bool leadingblanks; 
    leadingblanks=((c==blank)||(c==lf)||(c==cr));skipblanks();
    string word; word=emptyword; 
    while ((!inpstr.eof()) && (c!=blank) && (c!=lf) && (c!=cr)) 
        {word +=c; c=inpstr.get();}
    if (word.length()>0) 
       {if (leadingblanks) 
           {if (linelen+word.length()+1>MaxPos) 
              {linelen=word.length(); 
               outpstr << endl << word;}
            else  {linelen=linelen+word.length()+1;
               outpstr << blank << word;}}
        else {outpstr << word; linelen=word.length();}}}
\end{verbatim}
The competence domain ($CD_4$) of this program is the set of sequences which have no
words longer than MaxPos.
The fifth program takes into account the possibility of encountering long
words in the input stream, and proceeds to set the boolean flag {\tt longword}
to true, while skipping the long words.
\begin{verbatim}
p5:   //  same as p4, except:
int main ()
   {// ... ... ...
    c=inpstr.get();linelen=0;longword=false;skipblanks();...} 
void echoword ()
   {bool leadingblanks; 
    leadingblanks=((c==blank)||(c==lf)||(c==cr));skipblanks();
    string word; word=emptyword; 
    while ((!inpstr.eof())&&(c!=blank)&&(c!=lf)&&(c!=cr)) 
        {word +=c; c=inpstr.get();}
    if (word.length()>0) 
       {if (word.length()>MaxPos)
           {longword=true;}
        else {if (leadingblanks) 
              {if (linelen+word.length()+1>MaxPos) 
                  {linelen=word.length(); 
                   outpstr << endl << word;}
               else  {linelen=linelen+word.length()+1;
                   outpstr << blank << word;}}
           else {outpstr << word; linelen=word.length();}}}}
\end{verbatim}
The competence domain ($CD_5$) of this program is the set of all input sequences.

As we can see, $CD_0\subseteq CD_1\subseteq CD_2\subseteq CD_3\subseteq CD_4
\subseteq CD_5$, hence 
$P_0\refinedby_R P_1\refinedby_R P_2\refinedby_R P_3\refinedby_R P_4\refinedby_R 
P_5$.  Also, because $CD_5=dom(R)$, where $R$ is the (unwritten, but available
in \cite{mili}) specification, we
infer that $P_5$ is correct with respect to $R$.

We have developed our programs in a stepwise manner, by considering
broader and broader subsets of the domain of the specification, until
we reach the whole domain.  To some extent, the
transition from one
program to the next preserves much of the code that has been written, and
modifies/ adds relatively little code.
Perhaps more interestingly, the stepwise correctness enhancements
enable us to tackle the complexity of the specification one issue
at a time, and to validate our solution for one step before we
tackle the next step:  we have coded the
file processing aspects in {\tt p1}, the extra (leading, trailing,
middle) blanks in {\tt p2}, the removal of incoming newlines in
{\tt p3}, the insertion of outgoing newlines in {\tt p4}, and
the detection of long words in {\tt p5}.  At each step, we ensure
that the program works properly for the targeted competence domain
before we consider the next (broader) competence domain.

\section{Programming for Reliability}
\label{reliabilitysection}

The reliability of
a program $P$ can be defined with respect to two parameters: a
specification $R$ in the form of a binary relation; 
and a discrete probability distribution $\probdist$ over the domain of $R$, reflecting 
a given usage pattern.  We have seen in section \ref{relcorsect}
that for deterministic programs,
enhanced correctness logically implies (but is not equivalent to) 
enhanced reliability.  This means that if the derivation of a correct
program $P$ from a specification $R$ proceeds through a sequence of
increasingly correct programs, say $P_0$, $P_1$, $P_2$, etc.. $P_n=P$,
then the sequence of $P_i$'s is ordered by increasing reliability. So
that the only difference between deriving a correct program and deriving
a sufficiently reliable program (for a required reliability threshold) is
that in the latter case we can end the derivation earlier, namely as
soon as the reliability of $P_i$ matches or exceeds the selected threshold.
Given that correctness is the culmination of reliability, it is only
fitting that the derivation of correct programs be the culmination of
the derivation of reliable programs.

To illustrate our claim, we consider the four sample program derivations
presented in the previous section, and for each example we derive a test
driver and a (random) test data generator.  Then we apply each test driver
to the sequence of programs $\{P_i\}$ generated in the corresponding derivation.
This allows us to estimate the reliability of each program $P_i$ in each
example;  The table below shows how the reliability evolves as we proceed
from one program to the next; the first column shows the size of the
(random) test data used for each example.
\begin{center}
\begin{tabular}{|l|r|r|r|r|r|r|r|}
\hline
  & \shortstack[l]{Test Data\\Size} & $P_0$ & $P_1$ & $P_2$ & $P_3$ & $P_4$ & $P_5$\\
\hline
Fermat & 4000 &0.0000 & 0.2535 & 0.3445 & 1.0000 & &\\
\hline
Sqrt Ceiling & 4000 & 0.0000 &$\approx$ 0.0000 & 0.1020 & 1.0000 & & \\
\hline
String Analysis & 100 & 0.0000 &  0.0057 & 0.2790 & 0.2917 & 1.0000 & \\
\hline
Word Wrap & 3000 & 0.0000 & 0.0363 & 0.0873 & 0.1023 & 0.8990 & 1.0000 \\
\hline
\end{tabular}
\end{center}

\section{Critique}
\label{takingstocksection}

\subsection{Refinement vs Correctness Enhancement}

To elucidate the contrast between refinement and correctness
enhancement, we revisit the concept of {\em projection}, discussed in
section \ref{projectionsection}.  
This operator has many projection-like properties, hence its name,
including:
\begin{itemize}
\item Idempotence: $\projection_R(\projection_R(P))=\projection_R(P)$.
\item The projection of $P$ over $R$ is refined by $P$ and by $R$.
\item Program $P$ is correct with respect to $R$ if and only if 
$\projection_R(P)=R$.
\item Program $P'$ is more-correct than program $P$ with respect to $R$
if and only if the projection of $P'$ over $R$ refines the projection
of $P$ over $R$.
\end{itemize}
This last property is interesting because it highlights the contrast
between refinement and correctness enhancement.  Whereas refinement
mandates that we refine all of $P$, relative correctness mandates 
that we only refine the projection of $P$ over $R$, which is known
to be less-refined than $P$.

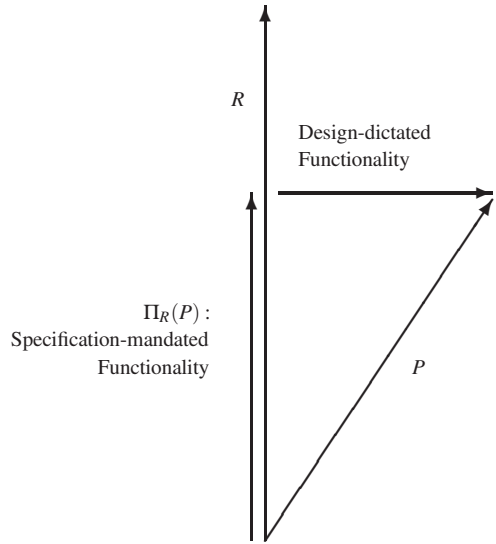
\begin{figure}
\thicklines
\setlength{\unitlength}{0.035in}%343in}
\begin{center}
\begin{picture}(100,80)

\put(50,0){\vector(0,1){80}}
\put(48,0){\vector(0,1){52}}
\put(52,52){\vector(1,0){32}}

\put(50,0){\vector(2,3){34}}

\put(45,65){{\scriptsize $R$}}
\put(55,56){{\scriptsize \shortstack[l]{Design-dictated\\Functionality}}}
\put(12,25){{\scriptsize \shortstack[r]{$\projection_R(P):$\\
Specification-mandated\\Functionality}}}
\put(72,25){{\scriptsize $P$}}

\end{picture}
\end{center}
\caption{\label{projectionfig}Decomposing the Function of a Program}
\end{figure}

The concept of projection enables us to distinguish between two
sources of functional properties in a program $P$:
\begin{itemize}
\item {\em Functional attributes that are mandated by the specification}.
This is the projection of $P$ over $R$.  
%In the case of the example cited
%in section \ref{projectionsection}, we have:
%\begin{itemize}
%\item $R=\{(s,s')| x'=x+y\}$,\\ on space $S$ defined by integer variables
%$x$ and $y$.
%\item {\tt p: while (y!=0) \{x=x+1;y=y-1;\}},\\ hence $P=\{(s,s')| y\geq 0
%\wedge x'=x+y\wedge y'=0\}$.
%\item $\projection_R(P)=\{(s,s')| y\geq 0\wedge x'=x+y\}$.
%\end{itemize}
\item {\em Functional attributes that are determined by design decisions}.
This the functionality that is delivered by $P$ but not mandated by $R$.
%In the case of the example cited in section \ref{projectionsection}, this
%is $\{(s,s')| y\geq 0\wedge y'=0\}$.  The specification does not ask for
%this functionality; rather this functionality stems from the design of $P$.
\end{itemize}
See Figure \ref{projectionfig}. The difference between refinement and 
relative correctness is that at each step, refinement refines all of $P$
whereas relative correctness refines only those functional attributes 
of $P$ that
are mandated by the specification; we may, in the process of doing so,
override design decisions made previously.  Because it makes no distinction
between specification-mandated attributes and design-dictated attributes
the paradigm of refinement must refine all the functional attributes of $P$;
consequently, every design decision taken during this process imposes
constraints on subsequent steps.

\subsection{Critique: Is Correctness Enhancement a Viable Methodology?}

Looking at the discussion of the previous section, one would be
forgiven for thinking that correctness enhancement is a panacea
for program derivation:
it involves refining a weaker specification than the traditional
refinement based process ($\projection_R(P)$
rather than $P$); we can take design decisions and not be constrained
by them (override them subsequently, if needed); we can stop halfway
through the derivation process and still have something to show for our effort
(an executable program that runs correctly for part of $dom(R)$); 
the artifacts we generatee as we proceed are increasingly reliable; etc.
But like everything else in life, this is too good to be true.  Indeed,
these advantages come at a cost, in terms of breadth of scope:  In
practice, correctness enhancement is viable only to the extent that
each artifact $P_i$ generated through this process can be derived
easily from the previous artifact, $P_{i-1}$.  But this is not always the
case:  just because $P_i$ represents a small increment of relative
correctness over $P_{i-1}$ does not necessarily mean that $P_i$
can be derived from $P_{i-1}$ by a simple syntactic modidfication.
In the examples of section \ref{samplesection} this was mostly true,
particularly for the word wrap example, where the stepwise correctness
enhancement process helped us deal with one aspect of the specification
at a time.  

Also, it is great that design decisions taken along the process do not
constrain subsequent design decisions, but this comes at the cost
of modularity:  in a refinement process, each design decision is validated
then fixed, and subsequent decisions are taken and validated accordingly.
But with correctness enhancement, at each step much of the new code must 
be analyzed and verified anew; there is some potential for verification
reuse, but it is not built into the process.
%But this is not always the case:  consider the space $S$ defined by an
%array {\tt a[0..N] of itemtype}, a variable $x$ of the type {\tt itemtype},
%and an index variable $k$: {\tt 0..N}, and let $R$ be the specification
%that provides for placing $0$ in $k$ if $x$ is not in $a[1..N]$, otherwise
%placing in $k$ a median index where $x$ occurs in $a[1..N]$.
%A possible sequence of programs may be:
%\begin{itemize}
%\item {\tt p0: \{abort;\}}.  Competence domain, $CD_0=\emptyset$.
%\item {\tt p1: \{k=0;\}}. $CD_1$: $x$ is not in $a[1..N]$.
%\item {\tt p2: \{k=N; a[0]=x; while (x=!a[k]) \{k=k-1;\}\}}.  $CD_2$:  
%$x$ does not appear more than twice in $a[1..N]$.
%\end{itemize}
%It is not clear how the next program, to treat the general case, can be
%derived as an amendment of {\tt p2}.  

Still, we argue that correctness enhancement is worthy of study, not
because it is better than refinement at the derivation of programs from
scratch, but rather because unlike refinement, correctness enhancement
models not only program derivation from scratch (a small and shrinking
segment of software engineering practice), but also the vast
majority of software engineering processes.
Indeed, we find that corrective maintenance, adaptive maintenance,
software merger, software upgrade, whitebox software reuse, 
extreme programming, and test
driven design are all instances of correctness enhancement.  
Much of software engineering practice consists, not of developing
a new product from scratch, but rather of taking a software product
that does not quite meet our needs, and evolving it to meet new 
requirements; this is essentially a correctness enhancement (with
respect to the new requirements) operation.

\subsection{Related Work}

To the extent that it can be seen as a weaker (less generic) form of
refinement, relative correctness bears some similarity with 
{\em retrenchment} \cite{banach}.  It is possible to think of 
relative correctness (more specifically: the property
of being less-correct than)
as being an instance of retrenchment, for a 
suitable choice of the {\em concedes} relation.
Yet, whereas retrenchment appears to apply to data abstraction in the context
of B specifications, relative correctness pertains primarily to programs
modeled
as mappings between initial states and final states.

Our notion of relative correctness is tightly coupled with our own
version of refinement, both in terms of its definition and in terms of
its notation.  We use relational algebra and we define refinement as:
$R'\refines R \Leftrightarrow (RL\cap R'L\cap (R\cup R')=R).$
While this formula captures, in relational terms, a fairly generic understanding
of refinement (larger domain, fewer images per argument), it is still
fairly different from other definitions of refinement, such as those
of \cite{abrial,back,hehner,hehnerec,hoare,morgan}.
An interesting venue of research would be to explore how to derive a
definition of relative correctness that corresponds to these
refinement formulas.

\section{Conclusion}
\label{conclusionsection}

Traditional refinement-based program derivation proceeds by 
successive correctness-preserving transformations starting from
the specification and ending with an executable program, when all
the specification notations have been replaced by program
statements.  In this paper we explore an orthogonal approach,
which starts from the trivially incorrect program {\tt abort} and
proceeds by successive correctness enhancing transformation until
we reach a correct program, or a sufficiently reliable program.  
While it offers many advantages, the proposed method is only applicable
to the extent that incremental correctness enhancements can be achieved
by commensurably incremental amendments to the text of the program;
this is a weighty caveat.

Regardless of its prowess in deriving programs from scratch,
and of its relative merit by comparison with refinement based programming,
correctness enhancement is worthy of study because it proves to be
an adequate model for a wide range of software engineering activities.
In order for these insights to be useful, we need to explore how 
mathematics of relative correctness can be turned into scalable
methods and tools, and how the results that we have derived for our
definition of refinement can be repurposed for other forms
of refinement.  This
is currently under investigation.

\subsubsection*{Acknowledgements}

The authors gratefully acknowledge the valuable feedback provided by the
anonymous reviewers.

\bibliographystyle{eptcs}
\bibliography{ref}

\begin{thebibliography}{10}
\providecommand{\bibitemdeclare}[2]{}
\providecommand{\surnamestart}{}
\providecommand{\surnameend}{}
\providecommand{\urlprefix}{Available at }
\providecommand{\url}[1]{\texttt{#1}}
\providecommand{\href}[2]{\texttt{#2}}
\providecommand{\urlalt}[2]{\href{#1}{#2}}
\providecommand{\doi}[1]{doi:\urlalt{http://dx.doi.org/#1}{#1}}
\providecommand{\bibinfo}[2]{#2}

\bibitemdeclare{book}{abrial}
\bibitem{abrial}
\bibinfo{author}{J.R. \surnamestart Abria\surnameend} (\bibinfo{year}{1996}):
  \emph{\bibinfo{title}{The B Book: Assigning Programs to Meanings}},
  \bibinfo{edition}{second} edition.
\newblock {\sl \bibinfo{series}{Series}} \bibinfo{volume}{443},
  \bibinfo{publisher}{Cambridge University Press}, \bibinfo{address}{Address},
  \doi{10.4204/EPTCS}.
\newblock \urlprefix\url{http://arxiv.org/abs/1009.3306}.
\newblock \bibinfo{note}{Note}.

\bibitemdeclare{book}{back}
\bibitem{back}
\bibinfo{author}{Ralph{-}Johan \surnamestart Back\surnameend} \&
  \bibinfo{author}{Joakim \surnamestart von Wright\surnameend}
  (\bibinfo{year}{1998}): \emph{\bibinfo{title}{Refinement Calculus - {A}
  Systematic Introduction}}.
\newblock \bibinfo{series}{Graduate Texts in Computer Science},
  \bibinfo{publisher}{Springer}, \doi{10.1007/978-1-4612-1674-2}.

\bibitemdeclare{inproceedings}{banach}
\bibitem{banach}
\bibinfo{author}{R.~\surnamestart Banach\surnameend} \&
  \bibinfo{author}{M.~\surnamestart Poppleton\surnameend}
  (\bibinfo{year}{2000}): \emph{\bibinfo{title}{Retrenchment, Refinement and
  Simulation}}.
\newblock In: {\sl \bibinfo{booktitle}{ZB: Formal Specifications and
  Development in Z and B}}, \bibinfo{series}{Lecture Notes in Computer
  Science}, \bibinfo{publisher}{Springer}, pp. \bibinfo{pages}{304--323},
  \doi{10.1007/3-540-44525-0\_18}.

\bibitemdeclare{book}{brink}
\bibitem{brink}
\bibinfo{author}{Chris \surnamestart Brink\surnameend},
  \bibinfo{author}{Wolfram \surnamestart Kahl\surnameend} \&
  \bibinfo{author}{Gunther \surnamestart Schmidt\surnameend}
  (\bibinfo{year}{1997}): \emph{\bibinfo{title}{Relational Methods in Computer
  Science}}.
\newblock \bibinfo{series}{Advances in Computer Science},
  \bibinfo{publisher}{Springer Verlag}, \bibinfo{address}{Berlin, Germany},
  \doi{10.1007/978-3-7091-6510-2}.

\bibitemdeclare{inproceedings}{ramicsdesh}
\bibitem{ramicsdesh}
\bibinfo{author}{J.~\surnamestart Desharnais\surnameend},
  \bibinfo{author}{N.~\surnamestart Diallo\surnameend},
  \bibinfo{author}{W.~\surnamestart Ghardallou\surnameend},
  \bibinfo{author}{M.~F. \surnamestart Frias\surnameend},
  \bibinfo{author}{A.~\surnamestart Jaoua\surnameend} \&
  \bibinfo{author}{A.~\surnamestart Mili\surnameend} (\bibinfo{year}{2015}):
  \emph{\bibinfo{title}{Relational Mathematics for Relative Correctness}}.
\newblock In: {\sl \bibinfo{booktitle}{RAMICS, 2015}}, {\sl
  \bibinfo{series}{LNCS}} \bibinfo{volume}{9348}, \bibinfo{publisher}{Springer
  Verlag}, \bibinfo{address}{Braga, Portugal}, pp. \bibinfo{pages}{191--208}.

\bibitemdeclare{article}{desharnais}
\bibitem{desharnais}
\bibinfo{author}{Jules \surnamestart Desharnais\surnameend},
  \bibinfo{author}{Nafi \surnamestart Diallo\surnameend},
  \bibinfo{author}{Wided \surnamestart Ghardallou\surnameend} \&
  \bibinfo{author}{Ali \surnamestart Mili\surnameend} (\bibinfo{year}{2017}):
  \emph{\bibinfo{title}{Projecting Programs on Specifications: Definitions and
  Implications}}.
\newblock {\sl \bibinfo{journal}{Science of Computer Programming}},
  \doi{10.1016/j.scico.2016.11.006}.

\bibitemdeclare{inproceedings}{oslo}
\bibitem{oslo}
\bibinfo{author}{Nafi \surnamestart Diallo\surnameend}, \bibinfo{author}{Wided
  \surnamestart Ghardallou\surnameend}, \bibinfo{author}{Jules \surnamestart
  Desharnais\surnameend} \& \bibinfo{author}{Ali \surnamestart Mili\surnameend}
  (\bibinfo{year}{2015}): \emph{\bibinfo{title}{Program Derivation by
  Correctness Enhacements}}.
\newblock In \bibinfo{editor}{John \surnamestart Derrick\surnameend},
  \bibinfo{editor}{Eerke~A. \surnamestart Boiten\surnameend} \&
  \bibinfo{editor}{Steve \surnamestart Reeves\surnameend}, editors: {\sl
  \bibinfo{booktitle}{Proceedings 17th International Workshop on Refinement,
  Refine@FM 2015, Oslo, Norway, 22nd June 2015.}}, {\sl
  \bibinfo{series}{{EPTCS}}} \bibinfo{volume}{209}, pp.
  \bibinfo{pages}{57--70}, \doi{10.4204/EPTCS.209.5}.

\bibitemdeclare{techreport}{dromey}
\bibitem{dromey}
\bibinfo{author}{Geoffrey \surnamestart Dromey\surnameend}
  (\bibinfo{year}{1983}): \emph{\bibinfo{title}{Program Development by
  Inductive Stepwise Refinement}}.
\newblock \bibinfo{type}{Technical Report} \bibinfo{number}{Working Paper
  83-11}, \bibinfo{institution}{University of Wollongong, Australia},
  \doi{10.1002/spe.4380150102}.

\bibitemdeclare{inproceedings}{gonzalezsanchez}
\bibitem{gonzalezsanchez}
\bibinfo{author}{Alberto \surnamestart Gonz{\'{a}}lez{-}Sanchez\surnameend},
  \bibinfo{author}{Rui \surnamestart Abreu\surnameend},
  \bibinfo{author}{Hans{-}Gerhard \surnamestart Gross\surnameend} \&
  \bibinfo{author}{Arjan J.~C. \surnamestart van Gemund\surnameend}
  (\bibinfo{year}{2011}): \emph{\bibinfo{title}{Prioritizing tests for fault
  localization through ambiguity group reduction}}.
\newblock In \bibinfo{editor}{Perry \surnamestart Alexander\surnameend},
  \bibinfo{editor}{Corina~S. \surnamestart Pasareanu\surnameend} \&
  \bibinfo{editor}{John~G. \surnamestart Hosking\surnameend}, editors: {\sl
  \bibinfo{booktitle}{26th {IEEE/ACM} International Conference on Automated
  Software Engineering {(ASE} 2011), Lawrence, KS, USA, November 6-10, 2011}},
  \bibinfo{publisher}{{IEEE} Computer Society}, pp. \bibinfo{pages}{83--92},
  \doi{10.1109/ASE.2011.6100153}.

\bibitemdeclare{book}{gries}
\bibitem{gries}
\bibinfo{author}{David \surnamestart Gries\surnameend} (\bibinfo{year}{1981}):
  \emph{\bibinfo{title}{The Science of Programming}}.
\newblock \bibinfo{publisher}{Springer Verlag},
  \doi{10.1007/978-1-4612-5983-1}.

\bibitemdeclare{inproceedings}{hehner}
\bibitem{hehner}
\bibinfo{author}{Eric C.~R. \surnamestart Hehner\surnameend} \&
  \bibinfo{author}{Andrew~M. \surnamestart Gravell\surnameend}
  (\bibinfo{year}{1999}): \emph{\bibinfo{title}{Refinement Semantics and Loop
  Rules}}.
\newblock In \bibinfo{editor}{E.D.I. \surnamestart Thor\surnameend} \&
  \bibinfo{editor}{E.~\surnamestart di~Thor\surnameend}, editors: {\sl
  \bibinfo{booktitle}{Formal Methods 1999}}, {\sl \bibinfo{series}{Lecture
  Notes in Computer Science}} \bibinfo{volume}{443},
  \bibinfo{organization}{Organization}, \bibinfo{publisher}{Springer Verlag},
  \bibinfo{address}{Address}, pp. \bibinfo{pages}{1--999}, \doi{10.4204/EPTCS}.
\newblock \urlprefix\url{http://arxiv.org/abs/1009.3306}.
\newblock \bibinfo{note}{Note}.

\bibitemdeclare{book}{hehnerec}
\bibitem{hehnerec}
\bibinfo{author}{Eric~C.R. \surnamestart Hehner\surnameend}
  (\bibinfo{year}{1992}): \emph{\bibinfo{title}{A Practical Theory of
  Programming}}.
\newblock \bibinfo{publisher}{Prentice Hall}, \doi{10.1007/978-1-4419-8596-5}.

\bibitemdeclare{incollection}{hoare}
\bibitem{hoare}
\bibinfo{author}{C.A.R. \surnamestart Hoare\surnameend} (\bibinfo{year}{1997}):
  \emph{\bibinfo{title}{Unified Theories of Programming}}.
\newblock In \bibinfo{editor}{E.D.I. \surnamestart Thor\surnameend} \&
  \bibinfo{editor}{E.~\surnamestart di~Thor\surnameend}, editors: {\sl
  \bibinfo{booktitle}{Mathematical Methods in Program Development}},
  \bibinfo{edition}{second} edition,
  \bibinfo{type}{type}~\bibinfo{chapter}{II}, {\sl \bibinfo{series}{Series}}
  \bibinfo{volume}{443}, \bibinfo{publisher}{Springer Verlag},
  \bibinfo{address}{Address}, pp. \bibinfo{pages}{1--999}, \doi{10.4204/EPTCS}.
\newblock \urlprefix\url{http://arxiv.org/abs/1009.3306}.
\newblock \bibinfo{note}{Note}.

\bibitemdeclare{book}{manna}
\bibitem{manna}
\bibinfo{author}{Zohar \surnamestart Manna\surnameend} (\bibinfo{year}{1974}):
  \emph{\bibinfo{title}{A Mathematical Theory of Computation}}.
\newblock \bibinfo{publisher}{McGraw-Hill}.

\bibitemdeclare{article}{meyer}
\bibitem{meyer}
\bibinfo{author}{Bertrand \surnamestart Meyer\surnameend}
  (\bibinfo{year}{1985}): \emph{\bibinfo{title}{On Formalism in
  Specification}}.
\newblock {\sl \bibinfo{journal}{IEEE Software}}
  \bibinfo{volume}{2}(\bibinfo{number}{1}), pp. \bibinfo{pages}{6--27},
  \doi{10.1109/MS.1985.229776}.

\bibitemdeclare{inproceedings}{ramics}
\bibitem{ramics}
\bibinfo{author}{A.~\surnamestart Mili\surnameend},
  \bibinfo{author}{M.~\surnamestart Frias\surnameend} \&
  \bibinfo{author}{A.~\surnamestart Jaoua\surnameend} (\bibinfo{year}{2014}):
  \emph{\bibinfo{title}{On Faults and Faulty Programs}}.
\newblock In \bibinfo{editor}{P.~\surnamestart Hoefner\surnameend},
  \bibinfo{editor}{P.~\surnamestart Jipsen\surnameend},
  \bibinfo{editor}{W.~\surnamestart Kahl\surnameend} \& \bibinfo{editor}{M.~E.
  \surnamestart Mueller\surnameend}, editors: {\sl
  \bibinfo{booktitle}{Proceedings, RAMICS 2014}}, {\sl \bibinfo{series}{LNCS}}
  \bibinfo{volume}{8428}, pp. \bibinfo{pages}{191--207}.

\bibitemdeclare{article}{mili}
\bibitem{mili}
\bibinfo{author}{A.~\surnamestart Mili\surnameend}, \bibinfo{author}{X.Y.
  \surnamestart Wang\surnameend} \& \bibinfo{author}{Y.~\surnamestart
  Qing\surnameend} (\bibinfo{year}{1986}): \emph{\bibinfo{title}{A Relational
  Specification Methodology}}.
\newblock {\sl \bibinfo{journal}{Software- Practice and Experience}}, pp.
  \bibinfo{pages}{1030--1030}.

\bibitemdeclare{book}{morgan}
\bibitem{morgan}
\bibinfo{author}{Carroll~C. \surnamestart Morgan\surnameend}
  (\bibinfo{year}{1998}): \emph{\bibinfo{title}{Programming from
  Specifications, Second Edition}}.
\newblock \bibinfo{series}{International Series in Computer Sciences},
  \bibinfo{publisher}{Prentice Hall}, \bibinfo{address}{London, UK}.

\bibitemdeclare{article}{miliscp}
\bibitem{miliscp}
\bibinfo{author}{Olfa \surnamestart Mraihi\surnameend}, \bibinfo{author}{Asma
  \surnamestart Louhichi\surnameend}, \bibinfo{author}{Lamia~Labed
  \surnamestart Jilani\surnameend}, \bibinfo{author}{Jules \surnamestart
  Desharnais\surnameend} \& \bibinfo{author}{Ali \surnamestart Mili\surnameend}
  (\bibinfo{year}{2013}): \emph{\bibinfo{title}{Invariant Assertions, Invariant
  Relations, and Invariant Functions}}.
\newblock {\sl \bibinfo{journal}{Science of Computer Programming}}
  \bibinfo{volume}{78}(\bibinfo{number}{9}), pp. \bibinfo{pages}{1212--1239},
  \doi{10.1016/j.scico.2012.05.006}.

\bibitemdeclare{inproceedings}{perelman}
\bibitem{perelman}
\bibinfo{author}{Daniel \surnamestart Perelman\surnameend},
  \bibinfo{author}{Sumit \surnamestart Gulwani\surnameend},
  \bibinfo{author}{Dan \surnamestart Grossman\surnameend} \&
  \bibinfo{author}{Peter \surnamestart Provost\surnameend}
  (\bibinfo{year}{2014}): \emph{\bibinfo{title}{Test Driven Synthesis}}.
\newblock In: {\sl \bibinfo{booktitle}{Proceedings, 35th ACM SIGPLAN
  Conference, PLDI}}, \bibinfo{volume}{49}, \bibinfo{address}{Edinburgh, UK},
  pp. \bibinfo{pages}{408--418}.

\bibitemdeclare{techreport}{reinfelds}
\bibitem{reinfelds}
\bibinfo{author}{Juris \surnamestart Reinfelds\surnameend}
  (\bibinfo{year}{1986}): \emph{\bibinfo{title}{A Brief Introduction to the
  Derivation of Programs}}.
\newblock \bibinfo{type}{Technical Report}, \bibinfo{institution}{University of
  Wollongong}, \bibinfo{address}{Wollongong, NSW Australia}.

\end{thebibliography}
%%\bibliography{refoxf}

\end{document}